\documentclass[12pt]{article}
\newcommand{\bq}{\begin{quote}}
\newcommand{\eq}{\end{quote}}
\newcommand{\be}{\begin{equation}}
\newcommand{\ee}{\end{equation}}
\newcommand{\ben}{\begin{enumerate}}
\newcommand{\een}{\end{enumerate}}
\newcommand{\bi}{\begin{itemize}}
\newcommand{\ei}{\end{itemize}}
\newcommand{\ket}[1]{\vert#1\rangle}

\newcommand{\braket}[2]{\langle#1\vert#2\rangle}
\newcommand{\ketbra}[2]{\vert#1\rangle\langle#2\vert}
\newcommand{\ts}[2]{\langle#1\Vert#2\rangle}
\begin{document}
\setlength{\baselineskip}{16.5pt}
\title{The Utility of Time-Symmetric Quantum Counterfactuals: A Response to Kastner%
\footnote{This paper was not accepted by \textit{Studies in History and Philosophy of Modern Physics} on the basis of a single reviewer's comments, reproduced here: ``Although this paper is neither bad nor obviously wrong, I see little reason for publishing it.  I can understand that an author who believes he has been misinterpreted in print feels an urge to set the record straight.  However, there can easily be no end to this process.  In my view, the main issue is whether publication will be of interest to potential readers.  Here the answer appears to be negative.  Sections 2 and 3 of the paper under review essentially summarize previous writings of Mohrhoff's.  Section 4 attempts to refute Kastner's objections.  Following the arguments developed in that section requires close familiarity with Kastner's paper as well as with others of both authors.  The point is that in order to understand Mohrhoff's ideas on TSQCs, it is much better and much easier to go to his original papers than to try to extract them from his reply to Kastner.  A version of the paper under review is posted on the arXiv web page.  I believe this is good enough to `set the record straight'.'' So do I, as a matter of fact.}}
\author{U. Mohrhoff\\
Sri Aurobindo International Centre of Education\\
Pondicherry 605002 India\\
\normalsize\tt ujm@auromail.net}
\date{}
\maketitle
\begin{abstract}\noindent Kastner's (Philosophy of Science 70, 2003, pp. 145--163) recent objections to the counterfactual usage of the time-symmetric Aharonov-Bergmann-Lebowitz rule by the author, especially her claims that the resulting time-symmetric quantum counterfactuals are vacuous or invalid, are examined and shown to be unfounded or beside the point.

\vspace{6pt}\noindent {\it Keywords\/}: ABL rule; time-symmetric quantum mechanics; quantum counterfactuals; Pondicherry interpretation

\setlength{\baselineskip}{14pt}
\end{abstract}
\section{Introduction}
Kastner (2003) has raised fresh objections to my counterfactual usage (Mohrhoff 2000) of the time-symmetric ABL rule (Aharonov \textit{et al.} 1964), adding to objections she raised in her (2001), most of which were met in my (2001). Here I shall show that these fresh objections, including those not adequately dealt with in my (2001), are either unfounded or beside the point. Section~2 stakes out the relevant semantic context. Section~3 introduces the specific question that time-symmetric quantum counterfactuals (TSQCs) address: how can we describe a quantum system between measurements? These two sections set the stage for the fourth, in which Kastner's objections are examined.

\section{The context}
\label{context}We are accustomed to the idea that the redness of a ripe tomato exists in our minds, rather than in the physical world---the world described by physical theory. We find it rather more difficult to accept that the same is true of the experiential now: it has no counterpart in the physical world. There simply is no objective way to characterize the present. The temporal modes past, present, and future can be characterized only by how they relate to us as conscious subjects: through memory, through the present-tense immediacy of sensory qualities, or through anticipation. In the physical world, we may qualify events or states of affairs as past, present, or future relative to other events or states of affairs, but we cannot speak of {\it the\/} past, {\it the\/} present, or {\it the\/} future. The proper view of physical reality is not only what Nagel (1986) has called ``the view from nowhere''---the physical world does not contain a preferred position corresponding to the spatial location whence I survey it---but also what Price (1996) has called ``the view from nowhen'': the physical world does not contain a preferred time corresponding to the particular moment (the present) at which I experience it. The idea that some things exist not yet and others exist no longer is as true (phenomenologically speaking) and as false (physically speaking) as the idea that a ripe tomato is red.

This is not a new insight. Augustine (1994) suggested long ago that time may be a dimension of the soul, rather than of the outer world.  In a letter of condolence to the sister and the son of his lifelong friend Michele Besso, Einstein wrote (Einstein and Besso, 1979): ``For us believing physicists, the distinction between past, present, and future is only a stubbornly persistent illusion.'' The fact that this distinction cannot be grasped by science, was to Einstein a matter of painful but inevitable resignation (Carnap, 1963). More recently the point was reiterated by Hans Primas and by Stephen Priest:

\bq All really fundamental physical dynamical laws are invariant under time translation and time reversal. Moreover, the concept of the ``now''---the brief interval that divides the past from the future---is absent in all fundamental mathematical formulations, both in classical physics and in quantum physics. That is, in a context-independent \textit{ontic} description \textit{there is no physical basis for the distinction between past and future}. (Primas, 2003, original emphases)

The tripartite temporal taxonomy has no physicalist or scientific or empirical explanation (never has, never will). Any explanation that is not empirical or scientific is metaphysical or theological so if the existence of past, present, and future can be explained, it can only be explained metaphysically or theologically. (Priest, 2006)

\eq As Primas points out, the fundamental physical laws are time-symmetric. They permit us to 
retrodict as well as to predict. Because the classical dynamical laws correlate events deterministically, they admit of causal interpretations. Or so it seems, for there is no physical basis for the associated causal \textit{arrow}.%
\footnote{Nor do we know anything about ``causal strings'' connecting the deterministically correlated states or events.}
It is we who project it into the physical world, based on our subjective and perhaps illusory sense of agency, which is made possible by a subjective temporal asymmetry: our ability to know the past as against our inability to know the future.

If, in addition, we project into the physical world the singular phenomenological presence of the present, we arrive at the flawed%
\footnote{If we imagine a spatiotemporal whole as a simultaneous spatial whole, then we cannot imagine this simultaneous spatial whole as persisting and the present as advancing through it. There is only one time, the fourth dimension of the spatiotemporal whole. There is not another time in which this spatiotemporal whole persists as a spatial whole and in which the present advances. If the experiential now is anywhere in the spatiotemporal whole, it is trivially and vacuously everywhere---or, rather, everywhen.}
conception of an evolving instantaneous state---a three-dimensional ``front'' advancing through four-dimensional spacetime. And if, in order to save a genuine (as against compatibilist) free will from the fatalism allegedly implied by the ``block universe'' of special relativity,%
\footnote{Stapp (2001), among others, has argued that the coexistence of the spatiotemporal whole implied by special relativity in turn implies that the future is as ``fixed and settled'' as the past. The coexistence of the spatiotemporal whole, however, is not simultaneous but tenseless and/or atemporal. It is a feature of the view from nowhen (Price, 1996). In this view time does not ``pass'' or ``flow,'' nor is there anything corresponding to the difference between the past and the future. An \textit{already} existing future would indeed by difficult to reconcile with the possibility of freely choosing a future course of action. But since an ``already existing future'' is a contradiction in terms, the only way to prevent the future from being determined by choices that are freely made \textit{now} is the possibility of \textit{foreknowledge}, and of this there is scant evidence.}
we embrace presentism, we arrive at the well-known folk tale according to which causal influences reach from the past to the future through persistent ``imprints'' on the present. A no-longer-existing past can influence a not-yet-existing future only through the mediation of something that persists. Causal influences reach from the past into the future by being ``carried through time'' by something that ``stays in the present.'' This evolving instantaneous state includes not only all presently possessed properties but also traces of everything in the past that is causally relevant to the future.

In classical physics, this is how we come to conceive of ``fields of force'' that evolve in time (and therefore, in a relativistic world, according to the principle of local action), and that mediate between the past and the future. The calculation of classical electromagnetic effects, for instance, can be carried out in two steps: given the distribution and motion of charges, we calculate a set of functions of position and time known as the ``electromagnetic field,'' and using these functions, we calculate the electromagnetic effects that these charges have on another charge. The rest is embroidery,%
\footnote{Even if it is considered ontology, it is embroidery in the sense that it adds nothing to our ability to predict or retrodict.}
viz., that the electromagnetic field is a physical entity in its own right, that it is locally generated by charges, that it mediates the action of charges on charges by locally acting on itself, and that it locally acts on charges.%
\footnote{As DeWitt and Graham (1971) so aptly put it, ``physicists are, at bottom, a naive breed, forever trying to come to terms with the `world out there' by methods which, however imaginative and refined, involve in essence the same element of contact as a well-placed kick.'' How does a charge locally act on the electromagnetic field? How does the electromagnetic field locally act on itself or on a charge? Apparently the familiarity of what seems to be local action induces us to believe that local action is self-explanatory.}

In quantum physics, this is why we tend to seize on algorithms that depend on the times of measurements (to the possible outcomes of which they serve to assign probabilities), to misconstrue the time dependence of such an algorithm as that of an evolving physical state, and to conceive of this state as mediating the dependence of the probabilities of possible outcomes on actual outcomes (or worse: as a state that determines the probabilities of possible outcomes without input from actual outcomes). This in turn accounts for our bafflement at the existence of EPR-type correlations (Einstein \textit{et al.}, 1935)---correlations between outcomes of measurements performed in spacelike relation that are impervious to both common-cause and mediatory explanations.

What is quantum mechanics sans embroidery? It is a mathematical formalism that serves to assign probabilities to possible measurement outcomes on the basis of actual outcomes. Anything else---any ``account of the nature of the external world and/or our epistemological relation to it that serves to \textit{explain} how it is that the statistical regularities predicted by the formalism with [this] minimal instrumentalist interpretation come out the way they do'' (Redhead 1987, 44, original emphasis)---is embroidery. This does not mean that everything else should be banned from physics, but it needs to be understood that anything else is pure, untestable metaphysics.%
\footnote{My views on what quantum mechanics is trying to tell us cannot, therefore, be filed away under ``instrumentalism'' or ``operationalism,'' nor would an attentive reader of my (2002, 2004b, 2005, or 2006) come away with such an impression. To say, as I do, that the irreducible core of quantum mechanics consists of algorithms that serve to assign probabilities to the possible outcomes of measurements on the basis of actual measurement outcomes is to state the obvious. From this irreducible core two lines of inquiry proceed. The one I consider fruitful analyzes the quantum-mechanical probability assignments in a variety of experimental contexts and arrives at a wide spectrum of ontological conclusions (Mohrhoff, 2002, 2004b, 2005, or 2006). The other, which I think is a red herring, aims to interpret some of the theory's mathematical symbols and/or equations as representing objective states and/or processes or posits underlying unobservables. Instead of addressing the important question as to \textit{why} measurements play a central r\^ole in what Redhead (1987, 44) called the ``minimal instrumentalist interpretation,'' it tries to sweep the question under the rug. As a referee (of a different paper) once put it to me, to solve the measurement problem ``means to design an interpretation in which measurement processes are not different in principle from ordinary physical interactions." I keep wondering what could possibly be meant by an ``ordinary physical interaction,'' considering that quantum mechanics describes interactions in terms of correlations between the probabilities of the possible outcomes of measurements performed on the interacting systems.}

\section{The question}
\label{question}According to Gisin (2002), the question ``How come the correlations?'' is ``one of the most---possibly the most---important for physics today.'' This question concerns the diachronic correlations between the outcomes of measurements performed on the same system at different times as much as the synchronic correlations between the outcomes of measurements performed in spacelike relation. We don't have more of an explanation for the diachronic correlations than we have for the synchronic ones. 

My question is less ambitious: how can we \textit{describe} a quantum system~$\cal S$ between consecutive measurements of two observables.

Suppose that measurements are repeatable, that the Hamiltonian is zero, and that the outcomes $a$ and~$b$ obtained at $t_a$ and~$t_b$, respectively, can be represented---for the purpose of calculating probabilities---by projectors into one-dimensional subspaces of the Hilbert space associated with~$\cal S$. Taking our cue from a time-honored sleight-of-hand---the transmogrification of a computational tool into a physical entity in its own right---we might interpret the ``retarded'' ket~$\ket a$ as representing the physical state of~$\cal S$ during the interval $[t_a,t_b)$. Or we might---with equal justification or lack thereof, given the time-symmetry emphasized in Sec.~\ref{context}---interpret the ``advanced'' ket~$\ket b$ as representing the physical state of~$\cal S$ during the interval $(t_a,t_b]$. Finally we might attempt to do justice to that symmetry by interpreting the time-symmetric ``two-state'' $\ts ab$ introduced by Aharonov and Vaidman (1991) as representing the physical state of~$\cal S$ during the interval $(t_a,t_b)$.

Rejecting this sleight-of-hand, we might posit retarded ``elements of reality'' corresponding to predictions of probability~1 (Einstein \textit{et al.}, 1935), advanced ``elements of reality'' corresponding to retrodictions of probability~1, or time-symmetric ``elements of reality'' corresponding to assignments of probability~1 based on the two-state $\ts ab$. The first would mean that $\cal S$ has property~$a$ during the entire interval $[t_a,t_b)$, the second would mean that $\cal S$ has property~$b$ during the entire interval $(t_a,t_b]$, and the last would imply that $\cal S$ has both properties during the entire interval $(t_a,t_b)$.

The problem with this strategy is that probability~1 is not sufficient for ``is'' or ``has.'' To see this, we only have to ask why the probability of finding a given particle in the union $A\cup B$ of disjoint regions $A,B$ equals the probability $p(A)$ of finding it in $A$ plus the probability $p(B)$ of finding it in~$B$. If $0 < p(A) < 1$, $0 < p(B) < 1$, and $p(A\cup B) = 1$, then it isn't certain that a perfect (100\% efficient) detector monitoring $A$ will click, it isn't certain that a perfect detector monitoring $B$ will click, yet it is certain that a perfect detector monitoring $A\cup B$ will click. How come? The answer is simplicity itself, if only we remember that quantum mechanics (sans embroidery) is a formalism that serves to assign probabilities to possible measurement outcomes on the basis of actual outcomes. Implicit in every quantum-mechanical probability assignment is the assumption that a measurement is successfully made: there is an outcome. So there is no mystery here, but it follows that quantum mechanics gives us probabilities with which this or that outcome is obtained in a \textit{successful} measurement. Probability~1 therefore means that a particular outcomes is certain given a successful measurement, not that a particular value is possessed regardless of measurements. One might postulate that it also means the latter, but this too would be untestable embroidery.

We are thus left with the possibility of describing a quantum system between measurements with the help of counterfactual probability assignments.%
\footnote{Whereas the probability distributions over the possible outcomes of all possible---but not actually performed---intermediate measurements contribute to describe, in terms of counterfactuals, a quantum system between measurements, the probability distributions belong to different possible worlds unless the outcomes over which they are distributed belong to compatible measurements. In addition one should distinguish between the counterfactual situation in which two compatible intermediate measurements are performed in a single possible world, and that in which the same measurements are performed in different possible worlds.}
All that is warranted is counterfactuals of the following type: if an observable~$\cal Q$ with nondegenerate eigenvectors $\ket{q_k}$ were measured during the interval $(t_a,t_b)$ given outcome~$a$ at~$t_a$, then the outcome represented by $\ketbra{q_k}{q_k}$ would be obtained with probability $p(q_k)=|\braket{q_k}a|^2$; if $\cal Q$ were measured during the same interval given outcome~$b$ at~$t_b$, then the outcome represented by $\ketbra{q_k}{q_k}$ would be obtained with probability $p(q_k)=|\braket b{q_k}|^2$; and if $\cal Q$ were measured during the same interval given both outcome~$a$ at~$t_a$ and outcome~$b$ at~$t_b$, then the outcome represented by $\ketbra{q_k}{q_k}$ would be obtained with probability
\be\label{ABL} 
p_{ABL}(q_k|a,b)=\frac{|\braket a{q_k}\braket{q_k}b|^2}{\sum_j|\braket a{q_j}\braket{q_j}b|^2}\,. 
\ee
While the first two probabilities are calculated with the help of the Born rule, the third is obtained by using the ABL rule. Just as the relative frequencies of outcomes obtained with ensembles that are preselected---the selection criterion being outcome $\ketbra aa$ at~$t_a$---tend to the probabilities $|\braket a{q_k}|^2$, so the relative frequencies of outcomes obtained with ensembles that are postselected---the selection criterion being outcome $\ketbra bb$ at~$t_b$---tend to the probabilities $|\braket b{q_k}|^2$. And, by the same token, the relative frequencies of outcomes obtained with ensembles that are pre- and postselected---the selection criterion being outcomes $\ketbra aa$ at~$t_a$ and $\ketbra bb$ at~$t_b$---tend to the probabilities $p_{ABL}(q_k|a,b)$. In the first case we discard all runs in which the outcome at~$t_a$ differs from $\ketbra aa$, in the second we discard all runs in which the outcome at~$t_b$ differs from $\ketbra bb$, and in the third we discard all runs in which the outcomes at~$t_a$ and~$t_b$ differ from $\ketbra aa$ and $\ketbra bb$, respectively.%
\footnote{At one point Kastner argued that the appropriate rule for conditional probability assignments to outcomes of measurements on pre- and post-selected systems is not the ABL rule but, instead, is given by equation (15) of her (1999b). In this equation the denominator includes interference terms that are absent from equation (\ref{ABL}) above. As I have shown in my (2001), Kastner's equation lacks self-consistency, inasmuch as in the numerator it assumes that the intermediate measurement is made, whereas in the denominator it assumes the contrary. This is not the way to do justice to the counterfactuality of TSQCs. Since counterfactuals are factual in a possible world, they ought to be calculated---both the numerator and the denominator---under the assumption that the intermediate measurement is made.}

The first to use the ABL rule counterfactually were Albert, Aharonov, and D'Amato (1985). A lively controversy ensued (Cohen, 1995; Kastner, 1999abc, 2001; Miller, 1996; Mohrhoff, 2001;  Sharp \&\ Shanks, 1993; Vaidman 1996ab, 1998ab, 1999abc). After the jointly published papers by Kastner (2001) and myself (Mohrhoff, 2001), during the preparation of which both authors were aware of their several times revised manuscripts, it seemed that the dust had settled, but two years later Kastner (2003) published a fresh critique of the use of time-symmetric counterfactuals by Vaidman%
\footnote{Vaidman has added to the tangle of ostensible and intended meanings by referring to certain time-symmetric counterfactuals as ``time-symmetric \textit{elements of reality}.''}
(1996ab, 1998ab, 1999abc) and myself (Mohrhoff, 2000, 2001).

\section{Kastner's issues}
\subsection{Irrelevance of the projection postulate}
Kastner (2003, 146) maintains that the ABL rule ``is essentially a time-symmetric generalization of the von Neumann Projection Postulate,'' and that it ``assumes that the density matrix of the system at the intermediate time~$t$ is a proper or `ignorance'--type mixture.'' As a matter of fact, the ABL rule is a time-symmetric generalization of the \textit{Born rule},%
\footnote{See my derivation of the ABL rule from the Born rule (Mohrhoff, 2001).}
which features in \textit{every} interpretation of quantum mechanics, whereas the projection postulate is another piece of untestable metaphysical embroidery, which can be seen from the fact that it only features in \textit{some} interpretations. The reason why the projection postulate does not appear in the ``Pondicherry interpretation of quantum mechanics'' or PIQM (Mohrhoff, 2000, 2004a, 2005), however, is the absence from this interpretation of an evolving quantum state rather than merely the absence of von Neumann's (1955) discontinuous mode of quantum state evolution. I fully agree with Peres (1984) that ``there is no interpolating wave function giving the `state of the system' between measurements.'' (According to the PIQM, the time dependence of probabilities is the dependence on the time of the measurement to the possible outcomes of which they are assigned, rather than the dependence on time of an evolving physical state.)

Does the ABL rule assume that the density matrix at the intermediate time~$t$ is a proper or `ignorance'--type mixture? Kastner persistently ignores the fact that implicit in every quantum-mechanical probability assignment are two assumptions: a measurement is made and an outcome is obtained. As said, quantum mechanics gives us probabilities with which this or that outcome is obtained in a \textit{successful} measurement. If I use either the Born rule or the ABL rule counterfactually, I must therefore assume that a measurement is made, even though in reality no measurement is made.  Needless to say that this is not an inconsistency but follows from the very nature of a counterfactual. So, indeed, in any possible world in which the intermediate measurement is performed (but its outcome is not taken into account) we have a proper or `ignorance'--type mixture. But only there. With respect to the real world this means: if the intermediate measurement \textit{were} performed, the density matrix of the system at the intermediate time~$t$ \textit{would} be a proper or `ignorance'--type mixture.

\subsection{No surprise}
Kastner (2003, 149) correctly takes the following formulation to be the intended meaning of my TSQCs:
\bi
\item[($1'$)]Consider system $\cal S$ having pre- and post-selection outcomes $a$ and $b$ at times $t_a$ and $t_b$ when a measurement of observable~$Q$ was not performed. If a measurement of observable~$Q$ had been performed on~$\cal S$ at time~$t$, $t_a < t < t_b$, and if $\cal S$ had the same pre- and post-selection outcomes as above, then outcome $q_j$ would have resulted with probability $p_{ABL}(q_j|a,b)$.
\ei
The corresponding possible-worlds formulation is
\bi
\item[(2)]In the possible world in which observable $Q$ is measured and system $\cal S$ yields outcomes $a$ and $b$ at times $t_a$ and~$t_b$, respectively, the probability of obtaining result $q_j$ at time~$t$ is given by $p_{ABL}(q_j|a,b)$.
\ei
In her (2003), Kastner has changed her view on these counterfactuals from ``inconsistent'' to ``trivial.'' She in fact believes that ($1'$) is as vacuous as the following:
\bi
\item[(A$'$)]If there had been a raffle at time~$t$, and if nobody had entered, then nobody would have won.
\ei
While (A$'$) is rather unsurprising, some TSQCs are, at least at first blush, quite surprising, what with Vaidman's (1996b) ``three boxes'' \textit{gedanken} experiment or my equivalent ``three holes'' experiment (Mohrhoff, 2001). The latter considers a particle launched at a specific location~$A$ in front of a plate with three holes%
\footnote{Whereas the initial and final states of the three-boxes experiment are superpositions of specific locations, the initial and final states of the three-holes experiment are not. This might have given the impression that the two experiments are not actually equivalent in all relevant respects. This, however, is a superficial impression, inasmuch as the relevant initial state is the prepared state of the particle as and when it reaches the plate containing the holes, while the relevant final state is the retropared state of the particle as and when it leaves the plate, and these are superpositions of specific locations.}
and warrants the following claims:
\bi
\item[(a)]If this particle is detected at a specific location~$B$ behind the plate, and if a certain measurement $M_1$ had been made, then one would have found with probability~1 that the particle went through the \textit{first} hole.
\item[(b)]If this particle is detected at~$B$, and if a certain measurement $M_2$ had been made instead, then one would have found with probability~1 that the particle went through the \textit{second} hole.
\ei
Kastner attempts to substantiate her claim that ($1'$) is as trivial as (A$'$) by arguing that the former is as unsurprising as the latter. If you want to ensure the truth of a claim as unlikely as ``if an attempt at holding a raffle at time~$t$ had been made, then nobody would have won,'' simply add the equally unlikely antecedent ``nobody entered.'' By the same token, if you want to ensure the truth of the unlikely claim ``if a particle launched at~$A$ in front of a plate with three holes had been subjected to a certain measurement, one would have found with probability~1 that the particle went through the first hole,'' simply add the unlikely antecedent ``the particle is subsequently detected at a specific location~$B$ behind the plate.'' Kastner claims that in both cases the element of surprise is put in ``by hand.''

This is a strange kind of argument. Quantum mechanics assigns probabilities to possible measurement outcomes on the basis of actual outcomes. We specify (i)~one or several actual outcomes and the respective times at which they are obtained, (ii)~a measurement $M$ with possible outcomes $m_k$, and (iii)~the time of~$M$. Quantum mechanics then gives us probabilities for the~$m_k$. \textit{Each} piece of information, on the basis of which the probabilities of the outcomes are calculated---via a state vector, a wave function, a density operator, or a two-state---is put in ``by hand.'' Of course it is unlikely that a particle launched at $A$ is found to have taken the first hole, if ``particle was launched at~$A$'' and ``$M_1$ was made'' is all the information provided. If the additional piece of information ``particle was detected at~$B$'' is provided, on the other hand, it is certain that the particle went through the first hole. The question of surprise does not arise. Recalling Popper's remark that our theories are ``nets designed by us to catch the world,'' Redhead concludes his (1987, 169) with the words: ``We had better face up to the fact that quantum mechanics has landed some pretty queer fish.'' If anything is surprising, it is quantum mechanics as a whole.

\subsection{A classical raffle}
Kastner not only claims that (1$'$) is as vacuous as (A$'$) but also that both counterfactuals are \textit{invalid}. According to her,
\bi
\item[(C)]a counterfactual is valid just in case
\bi\item[(i)]the antecedent nomologically implies the consequent and
\item[(ii)]the background conditions holding in the actual world ${\cal W}_a$ (where the antecedent is false) have no dependence on the truth value of the antecedent.
\ei\ei
As far as (A$'$) is concerned, Kastner makes the following identifications:
\bi
\item[(a)] ``there has been a raffle at time~$t$'' --- antecedent,
\item[(b)] ``nobody entered'' --- background condition,
\item[(c)] ``nobody won'' --- consequent.
\ei
She then argues that if an \textit{attempt} at holding a raffle is made, it is likely to succeed, and if it does, there are entrants. The only way we can force ``nobody won'' to nomologically follow from ``an attempt at holding a raffle was made'' is to stipulate that the background condition ``nobody entered'' does not change upon introduction of the antecedent. One wonders how this background condition can hold before the introduction of the antecedent ``there has been a raffle'' since it only makes sense in the context of a raffle. And if there has been a raffle, there were entrants---otherwise no raffle would have taken place. Kastner puts in the missing sense by changing the antecedent from (a)~``there has been a raffle'' to (a$'$)~``an attempt at holding a raffle was made.'' Such an attempt may occasionally fail, and so the consequent may occasionally be true, but of course it is not nomologically implied by the antecedent surreptitiously substituted for~(a).

Nor are we obliged to consider (b) a background condition. Whereas it doesn't make much sense to regard (a)\&(b) as a compound antecedent (which may lead one to think that (b) \textit{must} be considered a background condition), it is perfectly possible to regard (a$'$)\&(b) as a compound antecedent. This antecedent implies the consequent not just nomologically but logically. Moreover, there now isn't any background condition that could have a dependence on the truth value of the antecedent. Thus Kastner fails to establish the invalidity even of~(A$'$).

\subsection{Attack on a straw man}
Kastner (2003, 152) claims that the stipulation of the outcome at $t_b$ invokes 
\bq
a state of affairs that conflicts with the known processes of our world (such as: when raffles are held, people enter them; and when measurements are made at time~$t$, outcomes at time~$t_b$ generally don't occur with certainty but only with some probability dependent on the measurement outcome at time~$t$). 
\eq
What does Kastner believe the ``known processes of our world'' to be? Once again: the general theoretical framework of contemporary physics---quantum mechanics---is a probability calculus, a bunch of probability algorithms. In the specific context in which TSQCs are the \textit{only} tools to describe a system between consecutive measurements---the PIQM, which repudiates the notion that quantum ``states'' are evolving physical states (Mohrhoff, 2004a)---\textit{there are no known processes of the world}. All we have is correlation laws. It requires but a modicum of honesty to acknowledge that we don't know what is responsible for the validity and effectiveness of these laws. We don't know any process by which measurement outcomes influence the probabilities of measurement outcomes.

To the best of my knowledge nobody has ever claimed that when measurements are made at time~$t$, outcomes at time~$t_b>t$ generally occur with certainty. The passage just quoted is obviously designed to attack a straw man. If probabilities are assigned on the basis of all relevant \textit{earlier} outcomes, then a probability can be assigned to the possible outcome~$b$ at~$t_b$---in ${\cal W}_p$ this depends on the outcome at time~$t$ and generally differs from~1---but not to the outcome~$a$ at $t_a$ since this constitutes the assignment basis. If probabilities are assigned on the basis of all relevant \textit{later} outcomes, then a probability can be assigned to the possible outcome~$a$ at~$t_a$---in ${\cal W}_p$ this depends on the outcome at time~$t$ and generally differs from~1---but not to the outcome~$b$ at $t_b$ since this now constitutes the assignment basis. And if probabilities are assigned on the basis of all relevant earlier \textit{and} later outcomes, then a probability can be assigned neither to~$a$ at~$t_a$ nor to~$b$ at~$t_b$ since both outcomes then constitute the basis on which probabilities are assigned. All of these probability assignments yield valid counterfactuals, and none of them implies that the outcome at time~$t_b$ occurs with \textit{certainty}, either because the probability of this outcome generally differs from~1 or because this outcome is (part of) the basis on which probabilities are assigned.

\subsection{Inadequacy of philosophical analyses}
To date, the philosophical analysis of counterfactuals has almost exclusively been carried out in a framework that takes the ``flow'' of time for granted. This makes it virtually inapplicable to the issues at hand, as should be clear from Sec.~\ref{context}.

Goodman's (1947, 1983) seminal analysis, as noted by Kastner, was plagued by circularity: counterfactuals are defined in terms of cotenability, while cotenability is defined in terms of counterfactuals. The possible-worlds semantics of Lewis (2001) and Stalnaker (1984), which relies on a notion of closeness or similarity to the actual world, overcomes the circle in the truth-conditional schema of Goodman's metalinguistic approach, but it is plagued by the difficulty of defining and/or measuring similarity between worlds.

Lewis has the most similar world or worlds agree with the actual world right up to the time~$t$, at which a deviation from the laws of the actual world brings about the antecedent. From then on history again proceeds in accordance with the laws of the actual world. As D\"oring (1998) remarks, such an analysis precludes any counterfactual extrapolation of what the past would have had to have been in order to bring about the antecedent, even though  such ``backtracking'' is perfectly intelligible (and, I should add, neither vacuous nor invalid nor devoid of interest). Such an analysis is inapplicable to counterfactual probability assignments to earlier measurement outcomes based on later ones, including TSQCs, not only because of its inherent temporal asymmetry but also because no deviations from the quantum-mechanical correlation laws are required to bring about an antecedent that is false in the actual world.

Because no such deviations are required to bring about an antecedent that is false in the actual world, there is a unique and uniquely simple similarity criterion for TSQCs: that possible world is closest to the actual world which has \textit{all} the value-indicating events of the actual world \textit{plus one}---the measurement at time~$t$.

\subsection{A quantum raffle}
In my (2001) I pointed out a disanalogy between quantum and ``classical'' counterfactuals. In response to this, Kastner (2003, 155) discusses a  ``quantum raffle.'' Here is the gist of it, as far as I can tell:%
\footnote{Kastner's discussion is so muddled that it took me a long time to make sense of it.}
\bi
\item At time $t_a$, there are $N>0$ ``prospective entrants'' and an equal number of ``quantum coins.'' Each coin has a three-dimensional Hilbert space with a basis $\ket{\hbox{ready}}$, $\ket{\hbox{heads}}$, $\ket{\hbox{tails}}$ and is prepared in the ``ready state.'' (I would think of a ``ready state'' as the neutral state of a measuring device capable of indicating outcomes such as heads or tails, rather than as a three-state quantum system, but let this pass.)

\item If no raffle is held at time~$t$, each coin remains in the ready state. If a raffle is held at~$t$, each coin evolves into the ``flipped state,'' which is a superposition of heads and tails. (I would think of a ``flipped state'' as \textit{either} heads \textit{or} tails, but let this pass.) 

\item At time $t_b$, the number of entrants $H$ is determined by subjecting each coin to a measurement yielding either heads or tails. For each coin that comes up heads, a prospective entrant becomes an actual entrant.
\ei
Suppose that $H=0$, which implies that $T=N$, where $T$ is the number of tails found. There are then no actual entrants, no raffle was held, and each coin remained in the ready state. But if each coin remained in the ready state, the measurements at~$t_b$, designed to yield either heads or tails, will fail or be null, inasmuch as the ready state, being a member of a basis containing both heads and tails, is orthogonal to both heads and tails. In short, $T=N$ implies $T=0$. No such contradiction arises in the context of a TSQC.
 
Kastner in effect ties the possible outcomes of the measurement of ``raffle or no raffle'' to the state obtaining at~$t_b$ in such a way that inconsistencies are guaranteed. If no raffle took place, the initial, predictive state remains $\ket{\hbox{ready}}$ for all coins, in which case the final, retrodictive state cannot fixed at $\ket{\hbox{heads}}$ for some coins (or all coins, or none) and at $\ket{\hbox{tails}}$ for the remaining coins, since this would be inconsistent with the outcome ``no raffle.'' If there was a raffle, the final state is predicted to be $\ket{\hbox{heads}}$ for at least one coin and $\ket{\hbox{tails}}$ for the remaining coins, so that the final, retrodictive state cannot be fixed at $\ket{\hbox{ready}}$ for all coins, since this would be inconsistent with the outcome ``raffle.'' No such inconsistencies arise in the context of a TSQC.

Kastner (2003, 156) admits that ``the raffle differs from the usual TSQT [sic] in that there is a unitary evolution between~$t$ and~$t_b$ if the raffle is held'' but goes on to assert, incomprehensibly, that ``since such an evolution is fully time symmetric, the difference in no way disqualifies the example as a fair analogy.'' I fail to see how there can be a unitary evolution between~$t$ and~$t_b$ in one case and not in the other (whatever the other case may be). In which sense does she take unitary evolution to be time symmetric? We can ``evolve'' a state unitarily forward in time for the purpose of making predictions, and we can ``evolve'' a state unitarily backward in time for the purpose of making retrodictions. In this sense unitary evolution is time symmetric. But the state evolved forward from $t_a$ to~$t$ need not be the same as the state evolved backward from $t_b$ to~$t$. In this sense unitary evolution is not time symmetric. Finally, I fail to see how the full time symmetry of a unitary evolution implies that said difference does not disqualify Kastner's quantum raffle as a fair analogy.

\subsection{Classical counterfactuals vs. subjective ones}
Kastner (2003, original emphasis) attributes to both Vaidman and me the claim that ``certain `behind-the-scenes' features of quantum systems (i.e., questions of \textit{how it happens} that a system ends up with one outcome or another at times $t_a$ or~$t_b$)'' immunize TSQCs against comparisons with everyday counterfactuals. 

I won't vouch for Vaidman, but given my own publications on the subject this attribution is absurd. I have consistently rejected ``behind-the-scenes'' features of quantum systems; all we have to go by is measurement outcomes and their correlations. In keeping with this, I wrote (Mohrhoff, 2001, note~23):
\bq
While a classical counterfactual assumes that something obtains whereas in reality \textit{something else} obtains, a quantum counterfactual assumes that something obtains where in reality \textit{nothing} obtains.
\eq
Kastner (2003, 157) believes that, by this definition, what I have called the ``subjective counterfactual use of the ABL rule'' would constitute a classical counterfactual. To clarify, I have characterized probability assignments as ``subjective'' if and only if they contain an element of ignorance: they fail to take account of \textit{every} actual measurement outcome that has a bearing on (or makes a difference to) the probability assigned. What one might say is that a subjective counterfactual assignment using the ABL rule \textit{ignores} something that obtains---a relevant measurement outcome pertaining to the predictive state at~$t_a$ or to the retrodictive state at~$t_b$---while at the same time it assumes that something obtains (the measurement at the time~$t$) where in reality nothing obtains. What  I have called the ``subjective counterfactual use of the ABL rule'' therefore does \textit{not} constitute a classical counterfactual.

\subsection{``Counterfactual fixing''}
Kastner (2003, 158) maintains that ``Mohrhoff's tenseless view of facts---i.e., that a statement such as ``X~is true at time~$t_b$'' should be seen as holding at all other times---fails to accomplish the kind of counterfactual fixing he seeks.'' For the umpteenth time (in this and previously published papers), I do not seek any ``counterfactual fixing.'' What I seek is a way to describe quantum systems between measurements, and the only such way that does not introduce untestable metaphysics like probability algorithms transmogrified into evolving physical states, is to assign probabilities counterfactually and on the basis of all relevant outcomes past and future. 

``[I]f we are going to consider a counterfactual event at~$t$,'' so Kastner continues, ``then, to be consistent with physical law, we also have to consider possible outcomes at either $t_a$ or~$t_b$ other than the actual ones, that might have occurred but didn't.'' I fail to see the relevance of Kastner's reference to consistency with physical law. Nothing prevents us from considering situations in which the outcomes at $t_a$ or~$t_b$ are different from what they actually are. The probabilities of the possible outcomes of a not actually performed measurement at time~$t$ are as interesting given different outcomes at $t_a$ and/or~$t_b$ as they are given the actual outcomes. But they don't contribute to describe the system between $t_a$ and~$t_b$ given the actual outcomes at $t_a$ and~$t_b$, which I am concerned with.

\section{Conclusion}
In this paper Kastner's recent objections to my counterfactual usage of the time-symmetric ABL rule, especially her claim that the resulting time-symmetric quantum counterfactuals are vacuous or invalid, were examined and shown to be unfounded or beside the point.

\vspace{20pt}\setlength{\parskip}{5pt}\noindent\textbf{\large References}

\vspace{5pt}\leftskip=15pt\setlength{\parindent}{-15pt}
Aharonov, Y., Bergmann, P. G., \&\ Lebowitz, J. L. (1964). Time Symmetry in the Quantum Process of Measurement. \textit{Physical Review B}, 134, 1410--1416.

Aharonov, Y., \&\ Vaidman, L. (1991). Complete Description of a Quantum System at a Given Time. \textit{Journal of Physics A}, 24, 2315--2328. 

Albert, D. Z., Aharonov, Y., \&\ D'Amato, S. (1985). Curious New Statistical Predictions of Quantum Mechanics. \textit{Physical Review Letters}, 54, 5--7.
 
Augustine (1994). \textit{The Confessions of St. Augustine}, Bk. XI, Chap. XX, translated and edited
by A. C. Outler. New York: Dover.

Carnap, R. (1963). \textit{The Philosophy of Rudolf Carnap}, edited by Paul Arthur Schilpp, pp. 37--38. La Salle: Open Court.

Cohen, O. (1995). Reply to ``Validity of the Aharonov-Bergmann-Lebowitz Rule.'' \textit{Physical Review A}, 57, 2254--2255.

DeWitt, B. S., \&\ Graham, R. N. (1971). Resource letter IQM-1 on the interpretation of quantum mechanics. \textit{American Journal of Physics}, 39, 724--38.

D\"oring, F. (1998). Counterfactual Conditionals, in \textit{Routledge Encyclopedia of Philosophy}, Version 1.0 (Gen. Ed. Edward Craig). London and New York: Routledge.

Einstein, A., \&\ Besso, M. (1979). \textit{Correspondence 1903--1955}, Letter of March 21, 1955, p. 312. Paris: Hermann.

Einstein, A., Podolsky, B., \&\ Rosen, N. (1935). Can Quantum-Mechanical Description of Physical Reality be Considered Complete? \textit{Physical Review}, 47, 777--780.

Gisin, N. (2002). How come the correlations? Talk presented at the Symposium \textit{The Science of Nonlocality and Eastern Approaches to Exploring Ultimate Reality} organized by the Templeton Foundation, May 31, 2002.

Goodman, N. (1947). The Problem of Counterfactual Conditionals. \textit{Journal of Philosophy}, 44, 113--128.

Goodman, N. (1983). \textit{Fact, Fiction, and Forecast}. Cambridge, MA: Harvard University Press.

Kastner, R. E. (1999a). Time-Symmetrized Quantum Theory, Counterfactuals, and ``Advanced
Action.'' \textit{Studies in History and Philosophy of Modern Physics}, 30, 237--259.

Kastner, R. E. (1999b). The Three-Box Paradox and Other Reasons to Reject the
Counterfactual Usage of the ABL Rule. \textit{Foundations of Physics}, 29, 851--863.

Kastner, R. E. (1999c). TSQT ``Elements of Possibility''? \textit{Studies in History and Philosophy
of Modern Physics}, 30, 399--402.

Kastner, R. E. (2001). Comment on ``What Quantum Mechanics is Trying to Tell Us,'' by Ulrich Mohrhoff. \textit{American Journal of Physics}, 69, 860--863.

Kastner, R. E. (2003). The Nature of the Controversy over Time-Symmetric Quantum Counterfactuals. \textit{Philosophy of Science}, 145--163.

Lewis, D. K. (2001). \textit{Counterfactuals}. Malden, MA: Blackwell Publishers.

Miller, D. J. (1996). Realism and Time Symmetry in Quantum Mechanics. \textit{Physics Letters A}, 222, 31--36.

Mohrhoff, U. (2000). What Quantum Mechanics is Trying to Tell Us. \textit{American Journal of Physics}, 68, 728--745. 

Mohrhoff, U. (2001). Objective Probabilities, Quantum Counterfactuals, and the ABL Rule: A Response to R. E. Kastner. \textit{American Journal of Physics}, 69,  864--873.
 
Mohrhoff, U. (2002). Making Sense of a World of Clicks. \textit{Foundations of Physics}, 32, 1295--1311.

Mohrhoff, U. (2004a). Do Quantum States Evolve? Apropos of Marchildon's Remarks. \textit{Foundations of Physics}, 34, 75--97.

Mohrhoff, U. (2004b). This Elusive Objective Existence. \textit{International Journal of Quantum Information}, 2, 201--220.

Mohrhoff, U. (2005). The Pondicherry Interpretation of Quantum Mechanics: An Overview. \textit{Pramana---Journal of Physics}, 64, 171--185.

Mohrhoff, U. (2006). Is the End in Sight for Theoretical Pseudophysics? In \textit{New Topics in Quantum Physics Research}, edited by V. Krasnoholovets \&\ F. Columbus. New York: Nova Publishers.

Nagel, Th. (1986). \textit{The View from Nowhere}. New York: Oxford University Press.

Peres, A. (1984). What Is a State Vector? \textit{American Journal of Physics}, 52, 644--650.

Price, H. (1996). \textit{Time's Arrow \&\ Archimedes' Point}. New York: Oxford University Press.

Priest, S. (2006). Radical Internalism. \textit{Journal of Consciousness Studies}, 13, 147--174.

Primas, H. (2003). Time--Entanglement Between Mind and Matter. \textit{Mind and Matter}, 1,  81--119.

Redhead, M. (1987). \textit{Incompleteness, Nonlocality and Realism}. Oxford: Clarendon.

Sharp, W. D., \&\ Shanks, N. (1993). The Rise and Fall of Time-Symmetrized Quantum
Mechanics. \textit{Philosophy of Science}, 60, 488--499.

Stalnaker, R. C. (1984). \textit{Inquiry}. Cambridge, MA: MIT Press.

Stapp, H. P. (2001). Quantum Theory and the Role of Mind in Nature. \textit{Foundations of Physics}. 31, 1465--1499.

Vaidman, L. (1996a). Defending Time-Symmetrized Quantum Theory. arXiv: 9609007 [quant-ph].

Vaidman, L. (1996b). Weak-Measurement Elements of Reality. \textit{Foundations of Physics},
26, 895--906.

Vaidman, L. (1998a). On the Validity of the Aharonov-Bergmann-Lebowitz Rule. \textit{Physical
Review A}, 57, 2251--2253.

Vaidman, L. (1998b). Time-Symmetrical Quantum Theory. \textit{Fortschritte der Physik}, 46,
729--739.

Vaidman, L. (1999a). Defending Time-Symmetrized Quantum Counterfactuals. \textit{Studies
in History and Philosophy of Modern Physics}, 30, 373--397.

Vaidman, L. (1999b). Time-Symmetrized Counterfactuals in Quantum Theory. \textit{Foundations of Physics}, 29, 755--765.

Vaidman, L. (1999c). The Meaning of Elements of Reality and Quantum Counterfactuals---Reply to Kastner. \textit{Foundations of Physics}, 29, 865--876.

Von Neumann, J. (1955). \textit{Mathematical Foundations of Quantum Mechanics}. Princeton: Princeton University Press.

\end{document}